\documentclass[conference]{IEEEtran}
\IEEEoverridecommandlockouts
\usepackage{cite}
\usepackage{amsmath,amssymb,amsfonts}
\usepackage{algorithmic}
\usepackage{graphicx}
\usepackage{textcomp}
\usepackage{xcolor}
\usepackage{footnote}
\usepackage{braket}
\usepackage{hyperref}

\def\BibTeX{{\rm B\kern-.05em{\sc i\kern-.025em b}\kern-.08em
    T\kern-.1667em\lower.7ex\hbox{E}\kern-.125emX}}

\title{Quantum Neural Network for Accelerated Magnetic Resonance Imaging\\
\thanks{Shuo Zhou and Yihang Zhou contributed equally to this study and are the co-first authors. Haifeng Wang and Dong Liang are co-corresponding authors.

This work was partially supported by the National Key Technology Research and Development Program of China (2023YFB3811403,2023YFF0714201 and 2023YFC2411103), the National Natural Science Foundation of China (62271474, 62125111 and 52293425), the International Partnership Program of Chinese Academy of Sciences (321GJHZ2023246GC), the Guangdong Basic and Applied Basic Research Foundation (2024A1515012138), the High-level Talent Program in Pearl River Talent Plan of Guangdong Province (2019QN01Y986) and the Shenzhen Science and Technology Program (KJZD20230923113259001 and JCYJ20210324115810030).}
}

\author{
\IEEEauthorblockN{Shuo Zhou$^{\star}$ $^{\dagger}$,Yihang Zhou$^{\star}$ $^{\dagger}$,Congcong Liu$^{\star}$ $^{\dagger}$,Yanjie Zhu$^{\star}$ $^{\dagger}$,
Hairong Zheng$^{\star}$ $^{\dagger}$,Dong Liang$^{\star}$ $^{\dagger}$,Haifeng Wang$^{\star}$ $^{\dagger}$}
\IEEEauthorblockA{Email: s.zhou@siat.ac.cn yh.zhou2@siat.ac.cn cc.liu@siat.ac.cn yj.zhu@siat.ac.cn \\hr.zheng@siat.ac.cn dong.liang@siat.ac.cn hf.wang1@siat.ac.cn}
\IEEEauthorblockA{$^{\star}$Shenzhen Institute of Advanced Technology, Chinese Academy of Sciences, Shenzhen, China}
\IEEEauthorblockA{$^{\dagger}$University of Chinese Academy of Sciences, Beijing, China}
}

\begin{document}
\maketitle
\begin{abstract}
Magnetic resonance image reconstruction starting from undersampled k-space data requires the recovery of many potential nonlinear features, which is very difficult for algorithms to recover these features. In recent years, the development of quantum computing has discovered that quantum convolution can improve network accuracy, possibly due to potential quantum advantages. This article proposes a hybrid neural network containing quantum and classical networks for fast magnetic resonance imaging, and conducts experiments on a quantum computer simulation system. The experimental results indicate that the hybrid network has achieved excellent reconstruction results, and also confirm the feasibility of applying hybrid quantum-classical neural networks into the image reconstruction of rapid magnetic resonance imaging.
\end{abstract}


\section{Introduction}
Magnetic resonance imaging (MRI) is an indispensable tool for medical diagnosis and clinical research. However, due to the physical properties of MRI, it often requires a long scanning time to obtain clear images. The main methods for accelerating MR scanning include developing fast imaging sequences, hardware based parallel imaging, and reconstruction algorithms based on undersampling data. The reconstruction algorithm is based on signal processing methods to explore the prior information of MR images. Due to the emergence of compressive sensing (CS)\cite{b1}, sparse priors of images were applied to MR image reconstruction, and many other priors \cite{b2} were considered.

Since the emergence of deep learning, neural network has achieved excellent results in accelerating magnetic resonance reconstruction and become the mainstream method. Using neural network to learn the optimal parameters required for reconstruction from a large amount of training data or directly learn the mapping relationship between undersampling data and full sampling images \cite{b4}\cite{b5}, in order to achieve better imaging quality and higher acceleration than traditional parallel imaging or compressive sensing methods. Furthermore, with the development of deep learning and the improvement of network interpret ability, using prior information \cite{b6} and physical models \cite{b7}\cite{b8}  to design neural networks has achieved better results in magnetic resonance image reconstruction.

In recent years, with the development of quantum technology \cite{b9}, the field of quantum machine learning\cite{b10}\cite{b11} has also made rapid progress, and it has been found that quantum computing has many potential advantages in accelerating deep learning tasks. Large scale quantum computing is also difficult to truly implement due to hardware limitations. In order to solve this problem, hybrid algorithms of quantum networks and classical networks have emerged\cite{b12}\cite{b13}. Quantum computers essentially provide probabilistic results for the formation of coupled quantum systems during measurement, and due to their ability to perform large-scale parallel calculations on the superposition of quantum states, they can provide potential exponential acceleration\cite{b14}. And existing research has shown that quantum transformation of classical data can improve the accuracy of networks\cite{b15}\cite{b16}, which may be due to the potential ``quantum advantage" of quantum.

In this work, we designed a hybrid network of quantum network and classical network to learn end-to-end mapping of zero filled images and fully sampled images, and validated it on a publicly available dataset MoDL\cite{b17}. This network demonstrates good reconstruction results and also indicates that quantum network mixed with classical network have broad prospects in the field of medical imaging.
\begin{figure*}[htbp]
\centerline{\includegraphics[width=1\linewidth]{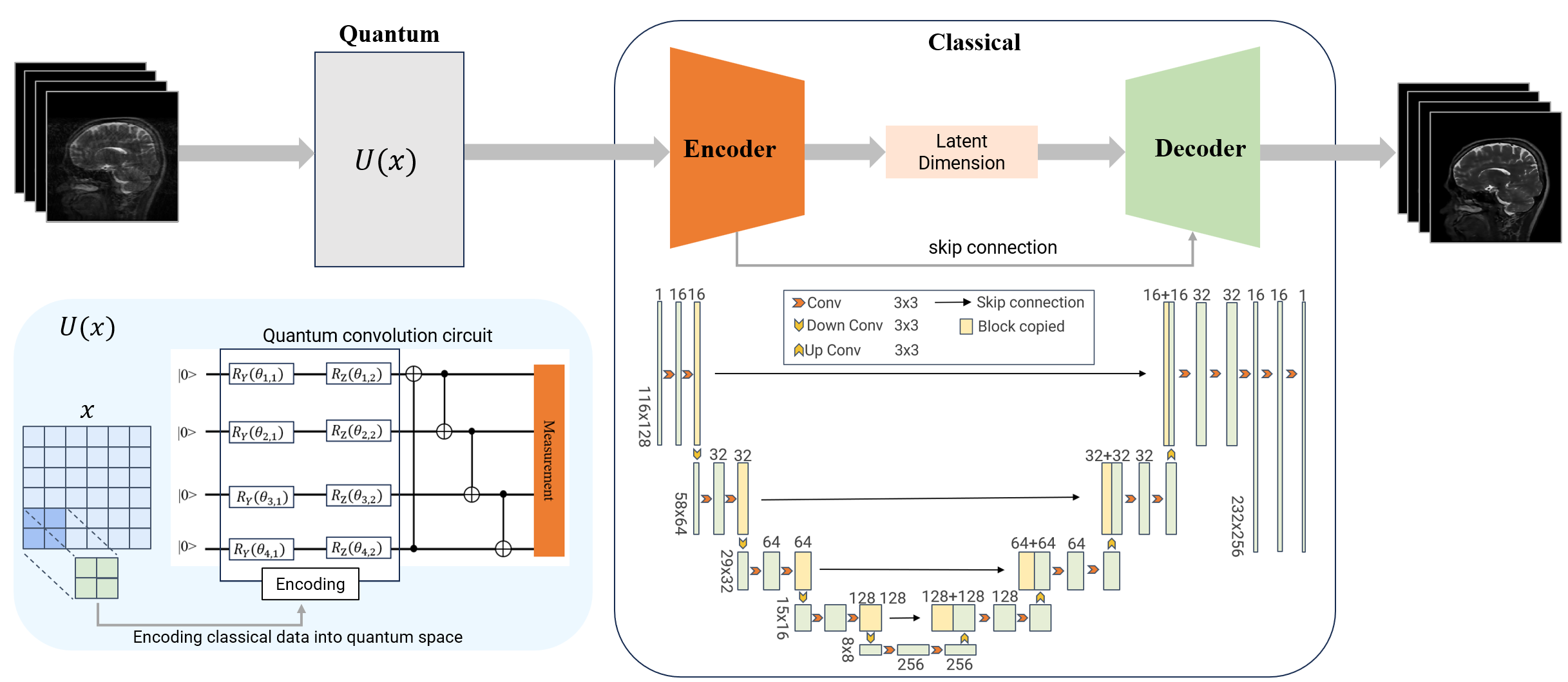}}
\caption{This is the structural diagram of a Hybrid Quantum Classic Neural Network, which includes the design of quantum circuits and the classical network structure of U-net.The top of the block displays the number of channels,the left side indicates the size of each layer.}
\label{fig1}
\end{figure*}
\section{Method}

In this study, a hybrid network architecture of quantum network and classical network was designed. The network learns the end-to-end mapping from zero filled images to fully sampled images, using zero filled images as data and fully sampled images as output. Figure 1 shows the detailed structure of the network, including the design of quantum circuits and the architecture of U-net network.

\subsection{Overall Training Formulation}

Let's represent the original (ground truth) image with y, and consider the undersampled original K-space data as

\begin{equation}
f=F_uy\label{eq1}
\end{equation}

$F_u$ represents the undersampled Fourier transform operator, $f$ represents the undersampled k-space data. The zero filled MR image z is generated as a direct inverse transformation of the observed k-space data

\begin{equation}
z=F_u^*F_u y\label{eq2}
\end{equation}

This article will learn an end-to-end quantum hybrid neural network to recover accurate MR images from zero filled data, attempting to reduce the following objectives

\begin{equation}
arg\min \limits_{\Theta}\{\sum\limits^T_{t=1}||C(z_t;\Theta)-y_t||_2^2 \}\label{eq3}
\end{equation}

$C$ is the end-to-end mapping function, $\Theta$ is the hidden parameter to be estimated, and $T$ is the total number of training samples.

\subsection{Quantum Convolution}

In the quantum convolution layer, we define the quantum convolution operation using $U(x)$. we designed a 4-bit quantum circuit to achieve quantum convolution with a kernel size of 2 × 2. Quantum entanglements are carried out using CNOT gates, and then the output of the quantum network is measured.

Quantum convolution kernels are the same as traditional convolution kernels that perform pixel by pixel convolution. The mainstream encoding methods for encoding classical data into quantum circuits currently include ground state encoding, angle encoding, and amplitude encoding\cite{b18}. In this article, angle encoding is selected, and based on the nature of the data and experimental comparison, the specific encoding method for angle encoding is

\begin{equation}
R_Y(\theta_{n,1})=R_Y(pix_n) \quad R_Z(\theta_{n,2})=R_Z(pix_n^2)\label{eq8}
\end{equation}

$R_Y$ and $R_Z$ are single qubit rotary gates, encoding classical data into quantum space. In each convolution operation, four pixels are selected, where n represents the index of the four pixels and $pix_n$ denotes the corresponding pixel value. According to the mapping relationship in Equation \ref{eq8}, classical data is encoded into the quantum circuit using $R_Y$ and $R_Z$ single qubit rotation gates.

We hope to use the random nonlinear features of quantum networks to help extract features that traditional networks cannot extract, thereby improving the accuracy of the entire network. Therefore, we have designed a quantum circuit to implement quantum convolution operations. In this work, a relatively simple implementation method was chosen, using only 2 × 2 quantum convolution kernels. For a 2 × 2 convolution kernel that requires 4 qubits, the size of the convolution kernel typically defines how many qubits a quantum circuit requires. For a quantum convolution kernel, we can express it in the following form

\begin{equation}
\ket{\psi}=\ket{\psi_1,\psi_2,\psi_3,\psi_4}\label{eq4}
\end{equation}

$\ket{\cdot}$ is the Dirac operator, representing the state of quantum bits, $\oplus$ represents the CNOT gate that is a double qubit gate. A quantum circuit is constructed to implement the above equation, as described in Figure 1. $R_Y$ and $R_Z$ are single qubit rotary gates, $\ket{0}$ represents the ground state of quantum and their initial states are all $\ket{0}$, then encode data into quantum bits through $R_Y$ and $R_Z$. $\ket{\psi_1}$-$\ket{\psi_4}$ are four qubits encoded with $R_Y$ and $R_Z$. The entangled CNOT gates are used to connect the four qubits for entanglement and stacking.

$\ket{\psi}$ represents the output result of a quantum convolution, and the use of quantum convolution kernels is similar to traditional convolution kernels. Pixels are convolved through sliding windows, and the output resolution of the convolution can also be set like traditional convolution operations.

$$
U(x)=Conv(x,\ket{\psi})
$$

Let $U(x)$ represent the result of the quantum convolution operation, where $x$ is the input image for the quantum convolution. The operator $\ket{\psi}$ denotes the quantum convolution kernel. $Conv$ represents the extraction of patches from 
$x$ by sliding window. These patches are then fed into the quantum circuit to perform the quantum convolution. This process is repeated until the entire image is processed, similar to the classical convolution.

This quantum circuit was built using the VQNet framework, which is a Python library developed by ORIGIN QUANTUM and executed on classical systems. Quantum circuit simulation is a key part of this library, which simulates quantum circuits on traditional hardware and tests quantum algorithms. In our research, quantum circuits were designed and simulated using VQNet, which is an important tool for preliminary verification of quantum machine learning techniques before actual deployment on quantum computing hardware. This experimental environment represents a general-purpose quantum computer capable of executing gate model instructions with arbitrary gate width, circuit depth, and fidelity. We will conduct experiments using noise free models and will conduct experiments on real quantum computers in the future.

\subsection{Classic Network}

The classical network part U-net has been chosen to be mixed with quantum networks.The U-net network consists primarily of an encoder, a decoder, and skip connections. The encoder downsamples the original input into a latent space, while the decoder upsamples it to progressively restore the image's spatial resolution. To preserve important features, skip connections are introduced to directly pass feature maps from earlier layers of the encoder to the corresponding layers of the decoder, thereby retaining high-resolution information and improving the network's accuracy. Although the network structure of U-net is relatively simple, existing research has shown that U-net also has good reconstruction effects on end-to-end magnetic resonance images \cite{b19}.

In order to achieve integration with quantum networks, this article has modified the traditional U-net. The network structure of U-net has become an asymmetric structure with two downsampling and three upsampling. Each downsampling reduces the resolution to half of the original, and each upsampling reduces the resolution to twice the original, depending on the input data. The number of channels in this U-net network varies from 1-16-32-64-128-256-128-64-32-16, which is an asymmetric structure, while the traditional U-net network structure is symmetric.Due to the quantum convolution layer reducing the image resolution to half of the input. To ensure that the initial input image and output image have the same resolution, we added an additional layer of upsampling at the end of the traditional U-net network for channel merging and resolution enhancement, as shown in block 3 in Figure 1.In the design of the network, all convolution kernels are set to 3 × 3, choose the default PyTorch parameter for the BatchNorm parameter and the activation function is chosen as ReLU.

\subsection{Combined Quantum Convolution and U-net}

The overall structure of the network is zero filled images passed through a quantum network, which performs the first layer of convolution and downsampling operations, and then enter the U-net network. The depth of the U-net network is asymmetric, and here it is mixed with the quantum network to ensure that the input and output dimensions of the entire network are equal.

The quantum convolution operation in this article is only achieved through the interaction between quantum particles without setting parameters, so the entire network only needs to train and learn U-net network parameters. The parameters that U-net needs to learn and the entire network have become as follows

\begin{equation}
    arg\min \limits_{\Theta}\{\sum\limits^T_{t=1}||C_{unet}(U(z_t);\Theta)-y_t||_2^2 \}\label{eq7}
\end{equation}

$C_{unet}$ represents the mapping of classical U-net network, with $\Theta$ as the parameter it needs to learn. The zero-filled MR image is denoted as $z$, $y$ represents the original (ground truth) image, $T$ is the total number of training samples and $U(\cdot)$ is the output result of the quantum convolution. This is an end-to-end training for quantum convolution output results to fully sampled images.

\section{Experiment}

\subsection{Comparison Experiment}

We chose end-to-end U-net as a comparative experiment. In order to ensure the rigor of the experiment and demonstrate the advantages of quantum properties, we added a convolution layer before U-net, whose convolution kernel size is 2 × 2, and the output resolution is half of the input, just like quantum convolution layer. Similarly, we also added an additional upsampling layer at the end of U-net to ensure that the resolution of the input and output images is the same. Overall, except for replacing quantum convolution layers with traditional convolution layers, all other parts are consistent with quantum hybrid neural networks.

\subsection{MR Image Reconstruction}

Data Set: The MRI data used public dataset MoDL\cite{b17} for this study were acquired using a 3D T2 CUBE sequence with Cartesian readouts using a 12-channel head coil. The matrix dimensions were 256 × 232 × 208 with 1 mm isotropic resolution. The training data had dimensions in rows × columns × slices × coils as 256 × 232 × 360 × 12 and testing data had dimensions 256 × 232 × 164 × 12.

A retrospective experiment was conducted using a random Cartesian sampling mask to simulate undersampling of fully sampled data to evaluate the performance of the model. Two times and four times acceleration experiments were conducted, respectively. The images were obtained by synthesizing multi coil data using the sum of square(SOS) method. The specific details are to use quantum convolution to downsample the zero filled image with an output resolution of half the original. Then, the result of quantum convolution is used as the input of the U-net network, and the fully sampled image is used as the output of the network for end-to-end training. The loss function is set to MSE. The testing data in the dataset consists of 164 images. In order to match the training data, we discarded some slice images of the brain edges and retained 90 images as the testing data.

\section{Result}
Reconstruction experiments were conducted with acceleration factors of 2 and 4, as illustrated in Figure 2 and 3. Separate networks were trained for each acceleration factor. For comparative purposes, U-net was used. Random Cartesian sampling with acceleration factors of 2 and 4 was applied for reconstruction. With an acceleration factor of 2, both U-net and the quantum hybrid neural networks demonstrated excellent reconstruction results. At an acceleration factor of 4, the quantum hybrid neural networks outperformed the traditional U-net in artifact reduction and image detail restoration, achieving superior reconstruction outcomes.

\begin{figure}[htbp]
\centerline{\includegraphics[width=1\linewidth]{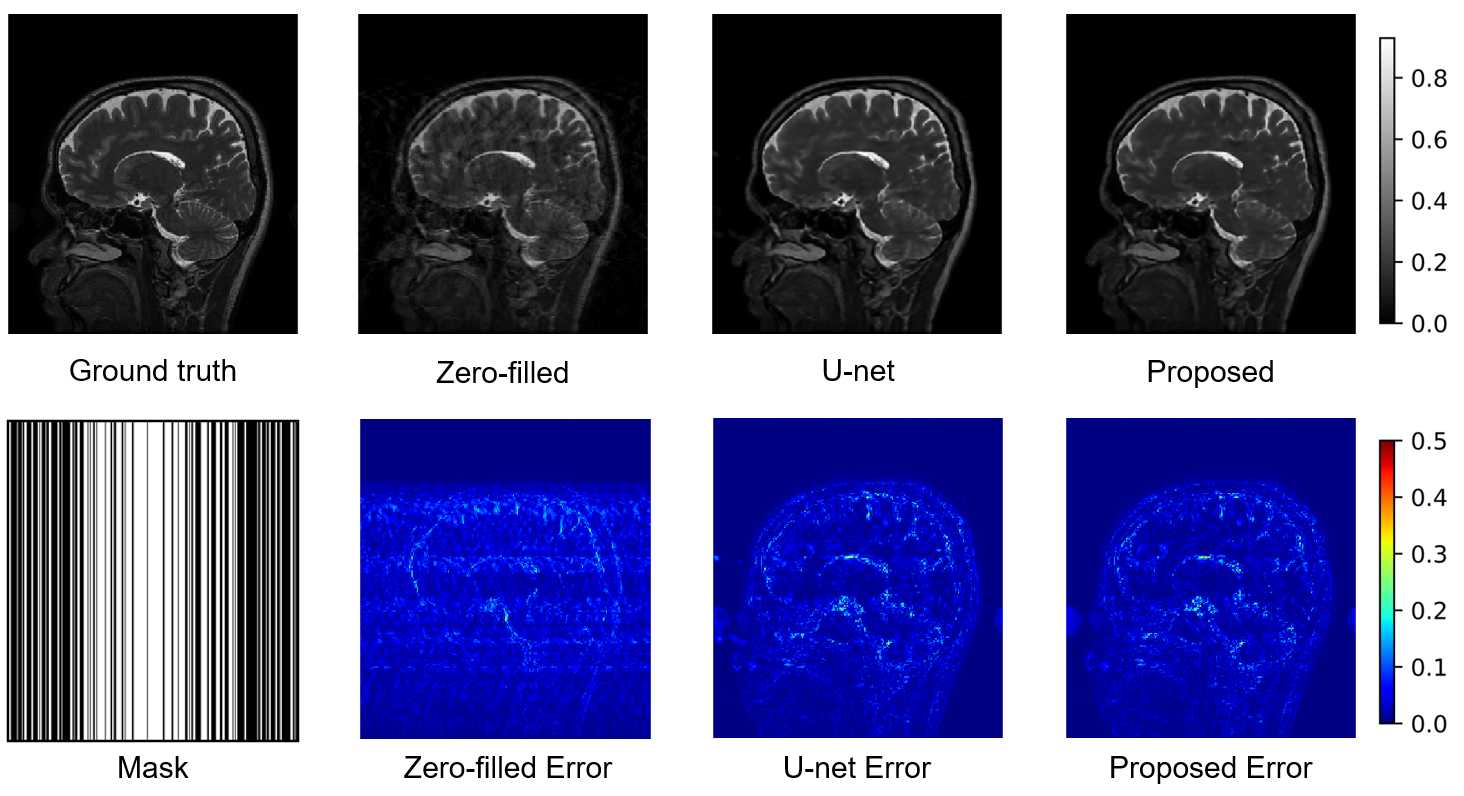}}
\caption{This image shows the reconstruction results with 2x acceleration. Includes the ground truth, mask, zero-filled image, U-net reconstruction result, quantum hybrid neural network reconstruction result, and their error map.}
\label{fig2}
\end{figure}

\begin{figure}[htbp]
\centerline{\includegraphics[width=1\linewidth]{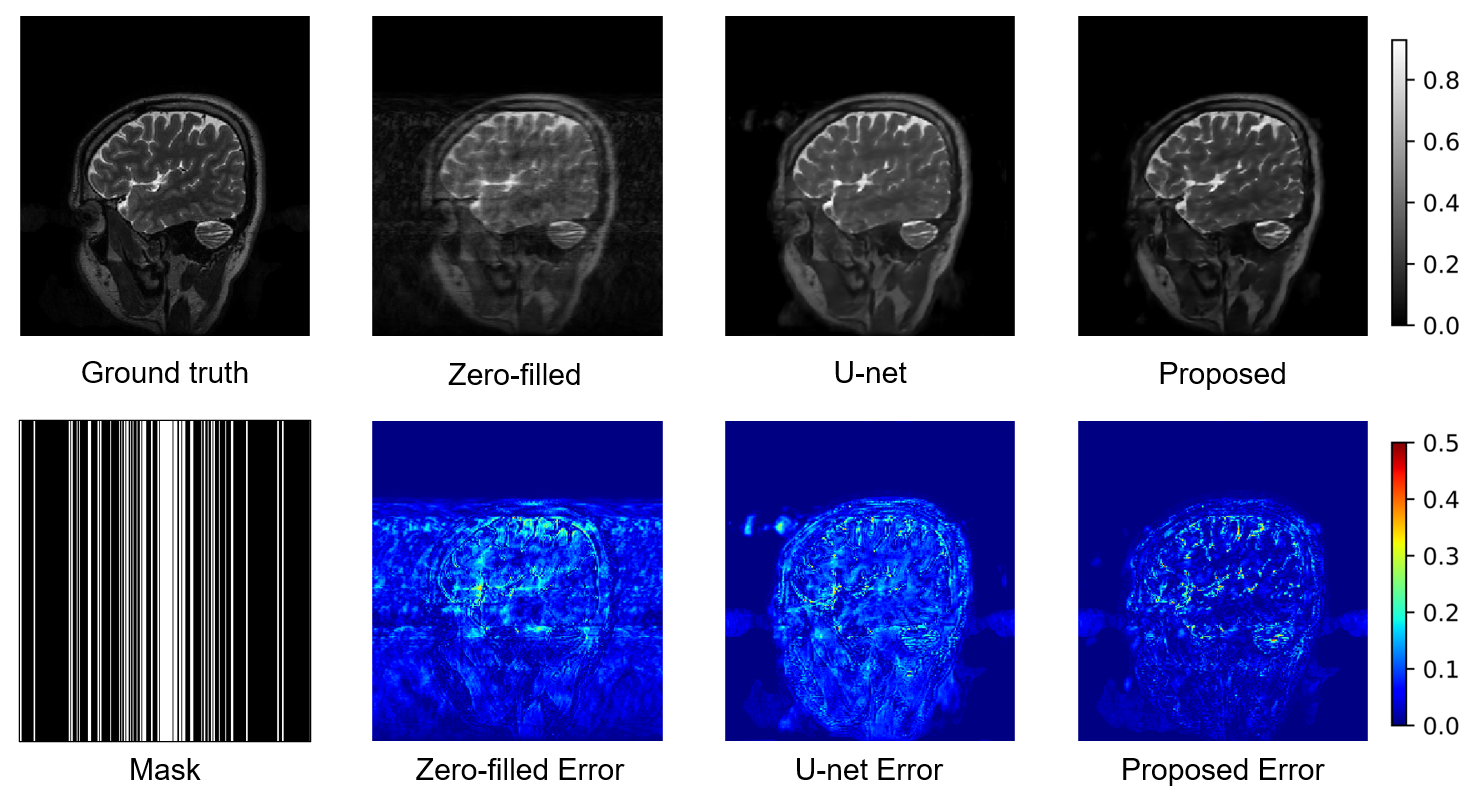}}
\caption{This image shows the reconstruction results with 4x acceleration. Includes the ground truth, mask, zero-filled image, U-net reconstruction result, quantum hybrid neural network reconstruction result, and their error map.}
\label{fig3}
\end{figure}
The quantitative evaluation of magnetic resonance reconstruction results was performed using Mean Squared Error (MSE), Peak Signal-to-Noise Ratio (PSNR), and Structural Similarity Index Measure (SSIM). Table \ref{tab1} presents the quantitative evaluation values for U-net and quantum hybrid neural network reconstruction results. A smaller MSE indicates that the reconstructed image is closer to the original image, with less deviation. Higher SSIM and PSNR values indicate better reconstruction results and a greater ability to restore the details of the original image.

\makesavenoteenv{tabular}

\begin{table}[htbp]
\centering
\caption{Quantitative indicators of reconstruction results}
\begin{center}
\begin{tabular}{lcccc}
\hline
& \textbf{U-net 2x} & \textbf{Proposed 2x} & \textbf{U-net 4x} & \textbf{Proposed 4x} \\
\hline
MSE  & 0.00093 & 0.00087& 0.00352  & 0.00323 \\
PSNR & 30.4671 & 30.7348 & 24.6199 & 25.1850 \\
SSIM & 0.8869  & 0.8906  & 0.7671  & 0.8251  \\
\hline
\end{tabular}
\label{tab1}
\end{center}
\end{table}

\begin{figure}[htbp]
\centerline{\includegraphics[width=1\linewidth]{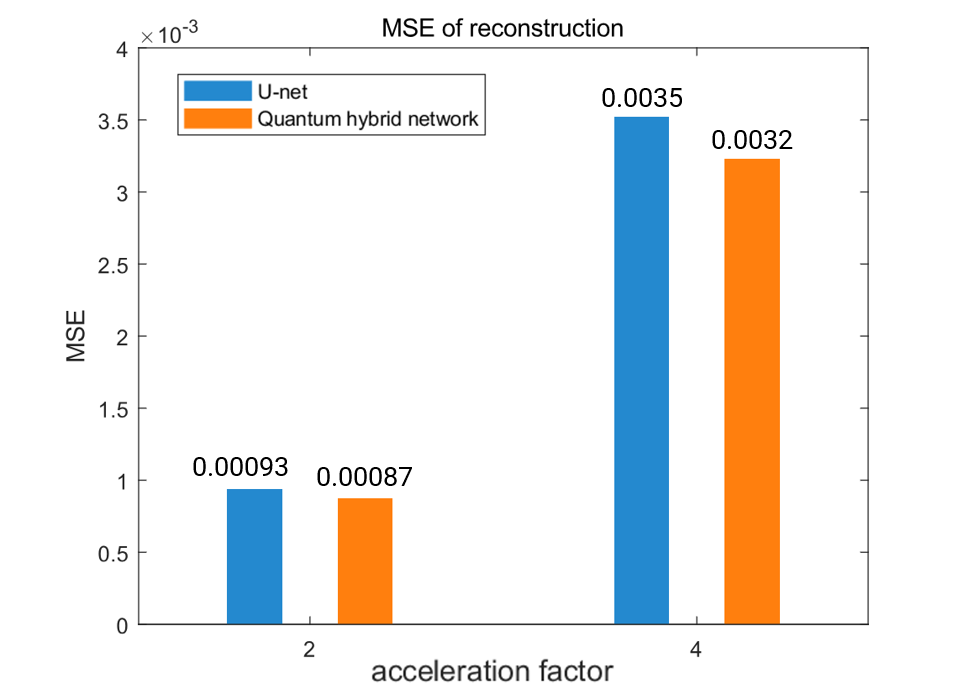}}
\caption{Bar chart of MSE for reconstruction results.}
\label{fig4}
\end{figure}

Figure 4-6 illustrate the bar chart comparing MSE, PSNR, and SSIM values. At an acceleration factor of 2, both the classical U-net and the proposed quantum hybrid neural network achieve excellent reconstruction results. Observations of the reconstructed images and corresponding quantitative metrics reveal comparable performance between the two methods. However, at an acceleration factor of 4, the quantum hybrid neural network outperforms classic network, demonstrating superior reconstruction with higher quantitative evaluation scores. This proposed method reflects much better reconstruction quality and structural similarity at the high acceleration factors than the traditional method.

The results indicate that the quantum hybrid neural network is both compatible and effective for magnetic resonance reconstruction. The enhanced accuracy may be attributed to potential quantum advantages. Future work will include further verification of these improvements on actual quantum computing hardware.

\begin{figure}[htbp]
\centerline{\includegraphics[width=1\linewidth]{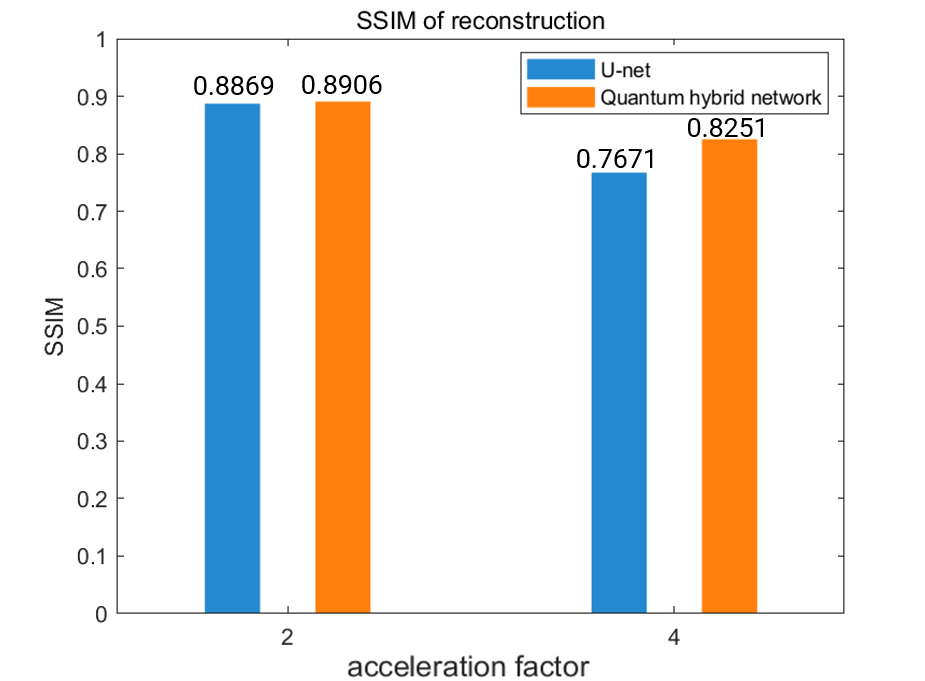}}
\caption{Bar chart of SSIM for reconstruction results.}
\label{fig5}
\end{figure}

\section{Conclusion}

This work attempts to combine quantum networks with traditional networks for end-to-end image reconstruction of rapid magnetic resonance imaging, verifying the feasibility of applying hybrid quantum-classical neural networks into magnetic resonance reconstruction. The experimental results of in vivo MR images show that
the proposed method can restore the lost fine structures and remove artifacts in zero filled MR images, demonstrating exciting reconstruction results. This fully demonstrates the potential and broad development prospects of quantum computing in the field of accelerating magnetic resonance imaging.

\begin{figure}[htbp]
\centerline{\includegraphics[width=1\linewidth]{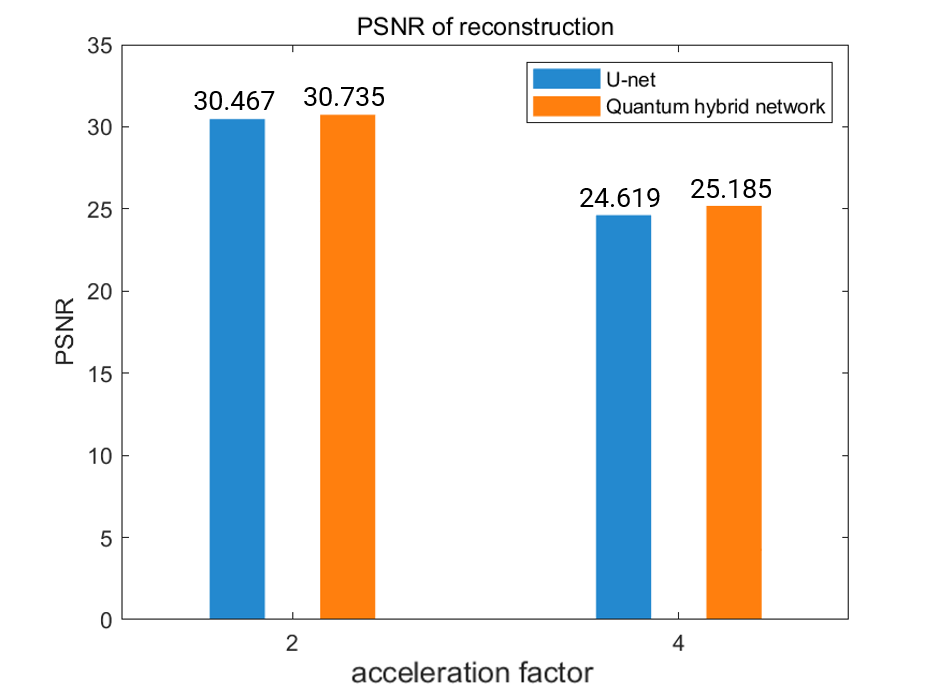}}
\caption{Bar chart of PSNR for reconstruction results.}
\label{fig6}
\end{figure}

\bibliographystyle{ieeetr}
\bibliography{ist.bib}

\end{document}